\documentclass[10pt,letterpaper]{article}
\usepackage[top=0.85in,left=1.in,footskip=0.75in,marginparwidth=2in]{geometry}

\usepackage[utf8]{inputenc}

\usepackage{cite}

\usepackage{nameref,hyperref}

\usepackage[right]{lineno}

\usepackage{microtype}
\DisableLigatures[f]{encoding = *, family = * }

\raggedright
\usepackage{changepage}

\usepackage[aboveskip=1pt,labelfont=bf,labelsep=period,singlelinecheck=off]{caption}

\makeatletter
\renewcommand{\@biblabel}[1]{\quad#1.}
\makeatother

\usepackage{lastpage,fancyhdr,graphicx}
\usepackage{epstopdf}
\fancyheadoffset[L]{2.25in}
\fancyfootoffset[L]{2.25in}

\usepackage{color}

\definecolor{Gray}{gray}{.25}

\usepackage{graphicx}

\usepackage{sidecap}

\usepackage{tabularx}
\usepackage{mathrsfs}
\usepackage{lineno}
\usepackage{epsfig}
\usepackage{amsmath}
\usepackage{amssymb}
\usepackage{float}
\usepackage{enumerate}
\usepackage{lineno}
\usepackage{booktabs}
\usepackage{arydshln}

\usepackage{wrapfig}
\usepackage[pscoord]{eso-pic}
\usepackage[fulladjust]{marginnote}
\reversemarginpar

\begin{document}
\vspace*{0.35in}

\begin{flushleft}
{\Large
\textbf\newline{Breakdown modes of capacitively coupled plasma I: transitions from glow discharge to multipactor}
}
\newline
\\
Hao Wu \textsuperscript{1}
Ran An\textsuperscript{1}
Dong Zhong\textsuperscript{1}
Wei Jiang\textsuperscript{2,*}
Ya Zhang\textsuperscript{3,*}
\\
\bigskip
\bf{1} School of Electronics and Information Engineering, Hubei University of Science and Technology, Xianning 437100, China
\\
\bf{2} School of Physics, Huazhong University of Science and Technology, Wuhan 430074, China
\\
\bf{3} Department of Physics, Wuhan University of Technology, Wuhan 430070,China
\\
\bigskip
* Corresponding Authors: weijiang@hust.edu.cn, yazhang@whut.edu.cn

\end{flushleft}

\section*{Abstract}
In this work, a one-dimensional direct implicit particle-in-cell/Monte Carlo collision (PIC/MCC) code is used to study the capacitive discharge driven under 60 MHz rf power in the background gas of pure argon. The electron-induced secondary electron emission (ESEE) model suitable for SiO$_2$ electrode is taken into account. Several discharge modes are found in the transition region between the higher-pressure glow discharge and the abnormal multipactor discharge in this simulation. In further, the abnormal multipactor will transform to normal multipactor with the contining decrease of pressure. These new discharges have higher electron energy and electron flux at the boundary and are mainly sustained by higher electrode-induced SEE coefficient and high frequency.

\section{Introduction}

 Plasma driven by radio frequency (rf) has been widely used in various industries, such as light and particle sources, and plasma jets. In the semiconductor industry, etching and deposition based on rf plasma play an irreplaceable role\cite{lieberman_principles_2005,chabert2011physics}. Driven by rf power, the capacitively coupled plasma (CCP) can be established and sustained easily and stably in a wide range of gas pressures from mTorr to Torr, in which the material can be processed under the controllable plasma parameters. In a steady state, both the gain and loss of energy and the number of particles (especially electrons) reach their balance\cite{chen2023note,wu2022note}. 
   
  In traditional stable discharge of CCP and similar plasma sources, for particle balance, the source of charged particles mainly comes from the ionization collisions in the discharge gap ($\alpha$ mode). For power balance, the energy of particles is mainly come from the electric field and is mainly lost through inelastic collision. In addition to gap ionization, electron emission from the surface of the electrode also contributes to the source of electrons under some circumstances.
  Secondary electron emission (SEE), as the most universal surface electron emission, can also provide a part of the electrons. SEE can be induced by electrons, ions, and neutral particles that bombard the electrode\cite{phelps1999cold,verboncoeur2005particle}. 

  In fact, only in an environment with a relatively stable electric field, such as a narrow gap, DC power, low-frequency rf power, in which heavy ions have enough relaxation time to move to the electrode with higher energy, ion-induced SEE (ISEE) can have a certain effect \cite{daksha2019material}. In DC-driven discharge, ISEE is an indispensable part for igniting and maintaining discharge ($\gamma$ mode)\cite{phelps1999cold,lafleur2013secondary}. 
  In the sheath-bulk-sheath structure of a stable rf plasma, the direction of the sheath electric field always points to the electrode surface, which will accelerate ion bombardment to the electrode, generating ion-induced SEE and affecting the discharge\cite{lafleur2013secondary}.
  
  For electron-induced SEE (ESEE), it plays a significant role in many devices, such as plasma thrusters \cite{sydorenko2006kinetic,kaganovich2007kinetic} and high-power microwaves \cite{kishek1998multipactor}. It also has an effect on the discharge of CCP to some extent\cite{horvath2017role,horvath2018effect}.
  In the initial time of CCP breakdown, SEE can supplement lost electrons, expanding the gas breakdown condition region and moving the breakdown curve to the lower pressure region\cite{vender1996simulations,smith2003breakdown,radmilovic2005modeling}. After the sheath is formed, the energy of the electron that bombards the electrodes is significantly reduced\cite{wu2021electrical}, so in most cases the ESEE is often ignored in the study of CCP. 
  When the gas pressure drops to an extremely low value, the glow discharge will not be able to establish and sustain the insufficient ionization collision rate \cite{lisovskiy1998rf}. 
  When a higher SEE coefficient is considered, under the force of the rf field, the electrons will accelerate to reach the electrode. In the case of high frequency, if the electrons arrive before the electric field changes its polarity, the secondary electrons emitted are repelled\cite{hohn1997transition}. Thus, the higher the driven rf frequency, the more conducive electron avalanche and gas breakdown. 
  
  In the case of frequency higher than 50 MHz, pressure lower than 5 mTorr, under the force of rf field, electron avalanche can also occur by ESEE from surface but ionization in the gap. Even if the sheath cannot be formed, the discharge can still be sustained. This unusual stable sustainable discharge mode is named multipactor\cite{hohn1997transition,kim2006transition}, which is commonly found in high-power microwave devices \cite{kishek1998multipactor}. 
  Multipactor discharge might break the vacuum gap in the field of ESEE and cause several deleterious effects on microwave components, such as rf electric noise, reflection of incident power, detuning of resonant structures, and production of heat\cite{hohn1997transition,udiljak2003new}. 
  
  For the plasma source, ESEE may be beneficial for discharge. However, in some microwave devices (micrometer to millimeter gap, gigahertz frequency, and surface dielectric films with high ESEE coefficient), the breakdown caused by surface electron multipactor discharges can easily occur and is extremely detrimental. Thus, a great deal of work has been applied to both the rf plasma and the multipactor, such as the effect of electrode shape\cite{na2019analysis},  
  single‐surface\cite{zhang2019analytical},
  magnetic\cite{hubble2017multipactor,spektor2018space}, backscattered electrons\cite{feldman2018effects},
  non-sinusoidal driven waveform\cite{iqbal2023two},
  harmonics\cite{iqbal2023recent,wen2022higher}.
  
  Compared to the frequencies of hundreds to several gigahertz in microwave devices, in several megahertz to tens of megahertz of rf-driven CCP, the multipactor is usually not obvious. Therefore, rarely have studies been applied to the characteristics of the multipactor in the CCP chamber.
  Fortunately, there are still several valuable references drawn up by earlier scholars.
  Through the rf discharge device and its various diagnostic equipment. F. H$\ddot{o}$hn and R. Beckmann \cite{hohn1997transition} draw out the region in which the gas can be broken down or the discharge can be sustained ( including the Paschen and extinction curve) in a wide range of gas pressure (about $10 ^{-3}\sim 10 ^{3}$ mTorr ) and rf voltage ( 10 V$\sim$2000V). 
  The transition from single-surface vacuum multipactor discharge to the rf plasma was investigated by H. C. Kim and J. P. Verboncoeur through the PIC/MC code \cite{kim2006transition}. The formation and difference of multipactor discharge and rf plasma have been analyzed in detail.
  
  Since glow discharge is useful for plasma sources but harmful for some microwave devices, exploring and studying the discharge boundaries as well as discharge connections is of great value for both the theory of gas discharge and the industry. By scanning the breakdown process of an rf-driven CCP similar device in a wide range of pressures and voltages through improved PIC/MCC code, the different sustainable discharge modes of the glow discharge and multipactor will be given and analyzed in more detail. 
  This may expand the application of rf discharge at extremely low pressure or provide a method for avoiding multipactor discharge damage to microwave devices. 
  Since several sustainable and unsustainable discharges were found in the simulation results, it is difficult to discuss them clearly in one paper. Therefore, the entire work is divided into two parts.
  As the first part of a two-part series, in this part, we will explore the sustainable discharge modes and their corresponding parameter spaces, which are organized as follows: the physical models and numerical methods are presented in Section 2. The simulation results of different stable discharge modes will be listed and analyzed in Section 3. Furthermore, the formation of different modes will be discussed in Section 4. Finally, conclusions will be drawn in Section 5.

\section{Computational model} \label{sec2}
  Two flat electrodes with a radius of $0.1$ m are placed symmetrically and opposite to each other to form the discharge gap, and their distance is set at 0.02 m. A 200 pF blocking capacitor is connected between the powered electrode (electrode1) and the rf power source, and another electrode(electrode2) is grounded. The RF power waveform satisfies $U_S(t) = U\sin{(2 \pi ft)}$, where $U$ and $f$ represent the voltage amplitude and frequency, respectively. $f$ is set to 60MHz.
  
  There will be a lot of electrons bombarding the electrodes during rf-driven discharge. If the energy of the primary electron is low, it might be absorbed or reflected. If the energy is greater than the emission threshold of the electrode surface (a dozen electron volts), the electron might be reflected, or absorbed, and a new secondary electron might be induced and emitted from the surface of the electrode to the gap. Therefore, the coefficient of ESEE ($\delta$) should be a function of the incident electron energy and is affected by the type of electrode material and the roughness of the surface\cite{verboncoeur2005particle,horvath2017role,horvath2018effect}. 
  In this work, the ESEE mode summarized by Horvath\cite{horvath2017role, horvath2018effect} is used to depict secondary electron emission, which is suitable for flat SiOelectrodes $_2$. The real SEE, elastically reflected electrons, and inelastically reflected electrons have been treated separately in Horvath's work.   In aluminum electrodes, the aluminum oxide film on the surface will make the emission coefficient greater than that of SiO$_2$ \cite{guo2019secondary}. In fact, in the high incident electron energy region (several hundred electron volts), the trend of the emission function is similar in different SEE modes \cite{vender1996simulations,smith2003breakdown}. 
  
  PIC/MCC has been widely used in simulating rf plasma and multipactor \cite{kim2006transition,na2019analysis}.
  In this work, we used a one-dimensional direct implicit particle collision in cell / Monte Carlo (PIC / MCC) coupled with an external circuit, which has been used successfully in previous work \cite{wu2021electrical,wu2022breakdown}. 
  The time step and the number of the grid are set to $5.0\times 10^-11$ and 65 respectively. 
  Since the implicit code allows for a larger time and space scale than the explicit one\cite{vahedi1993capacitive,kawamura2000physical,wang2010implicit}, this can finish the simulation more quickly.

   Pure argon is used as the background gas, and only the electron and Ar$^+$ are traced in this simulation. The consumption of the background gas is ignored since the ionization rate is extremely low in rf-driven low-pressure discharge.
   The MCC model proposed by Vahedi is adopted to deal with collisions \cite{vahedi1995monte}. The cross-sectional data come from \cite{phelps1999cold}. For electron-argon (e-Ar) collisions, elastic scattering, excitation, and ionization collisions are considered. For the ion-argon(e-Ar$^+$) collision, only charge exchange and elastic scattering are considered. 

   In this work, two simulating methods will be used. To explore the fast formation of an rf plasma or multipactor, the whole discharge process from extremely low electron density ($10^{8}m^{-3}$) to the last stable discharge state will be "diagnosed" and drawn. The final stable discharge is often the working state in most plasma sources, and it is also the state that can be easily diagnosed through experiments. To better know the steady state of discharge, more diagnostic codes will be added and run in more repeated periods to obtain the convergence result as accurately as possible.

\section{Results} \label{sec3}
\subsection{General consideration}
\begin{figure}[ht]
       \centering
         \includegraphics[width=0.8\textwidth]{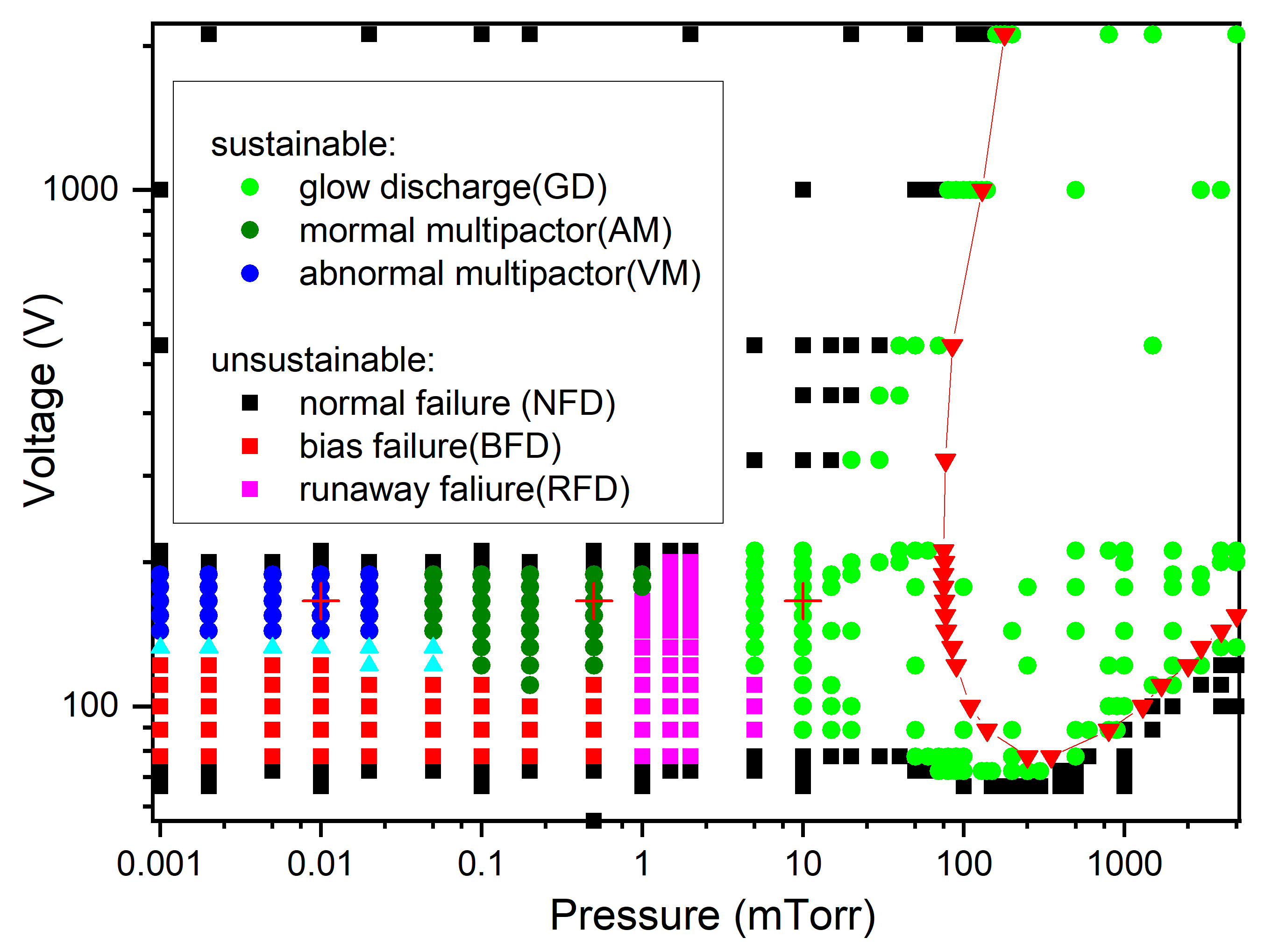}
         \caption{Different sustainable discharge modes zone in the frequency of 60MHz and distance of 2cm. The syan triangle represents the transition zone. The red triangle with red line is the breakdown curve without SEE. The points marked with red crosses represent the representative parameters of different sustainable discharge mode.}
      \label{BreakdownCondition}
    \end{figure}

    After a wide range scan of voltages and pressures, several discharge modes are found in the low-pressure region, as shown in figure \ref{BreakdownCondition}.
    The zone covered by circles with different colors (as shown in blue, olive, and green circles) represents the different types of sustainable discharge, which will be analyzed in detail in the following chapter; The zone covered by red, black, and magenta squares in figure \ref{BreakdownCondition} are the discharges that cannot be sustained including the normal failure discharge (BFD), bias failure discharge (BFD), runaway failure discharge (RFD) which means the particle density will go down or even disappear. The unsustainable discharge will be analyzed in Part II of the next paper. The cyan triangle region is the unstable transition zone, the discharge will not collapse, but it cannot be stabilized. 
    
   To better describe them, we renamed each sustainable mode and gave the representative points (discharge parameters with representative discharge characteristics) of them, shown in Table.\ref{SustainableDischargeName}. 
  \begin{table}[h]
        \centering
        \caption{Different sustainable discharge modes in 60MHz 2cm}
        \begin{tabular}{ccc}\hline
           Discharge mode  &  Abbreviate   &  Representative Point\\\hline
           glow discharge   &  GD  & 160V 10mTorr \\
           abnormal multapactor &  AM  & 160V 0.5mTorr \\
           normal multapactor & NM & 160V 0.01mTorr \\
           vacuum multapactor & VM & 160V Vacuum\\
           \hline
        \end{tabular}
        \label{SustainableDischargeName}
    \end{table}
    According to the existence of plasma, those sustainable discharge modes in low pressure can be divided into glow discharge (GD) and multipactor approximately. From the characteristics of the discharge, multipactor can be divided into abnormal multipactor (AM), and normal multipactor (NM), and vacuum multipactor (VM) in more detail. 
    
    GD is the traditional discharge of rf-driven CCP, which occupies a vast region in the upper right corner of the figure \ref{BreakdownCondition}.
    The multipactor is distributed in a narrow voltage zone of the extremely low-pressure region (blue and olive circle in figure \ref{BreakdownCondition}). 
    The time-averaged distribution of these sustainable discharges'  parameters is shown in figure \ref{StableNeiEeiPhi}. The different terms of flux and energy about electrons and ions at the surface of the powered electrode are shown in Table. \ref{ElectrodeParameterTable}. 
    Combined with those parameters, the characteristics of those discharge modes will be depicted more clearly respectively:
    
     \begin{figure}[ht]
       \centering
         \includegraphics[width=\textwidth]{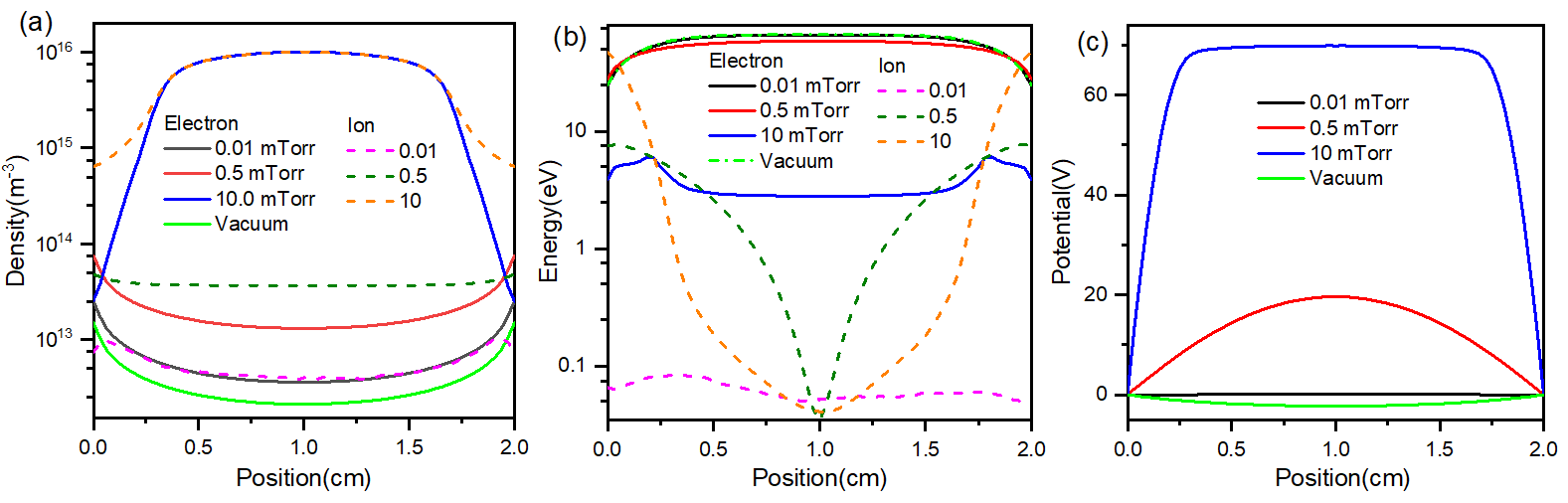}
         \caption{Time averaged parameter distribution at the steady-state in different pressure: (a) electron and ion density, (b) electron and ion energy,(c) potential. The pressure of 10mTorr, 0.5mTorr, and 0.01mTorr can represent the discharge type of GD, AM and NM}
      \label{StableNeiEeiPhi}
    \end{figure}

    \begin{table}[h]
    \centering
    \caption{Comparing of particle parameters at the electrode}
      \begin{tabular}{|c|c|c|c|c|} \hline
      \hline
       Discharge mode & GD & AM & NM & VM\\ \hline
       (V,P) & 160V 10mTorr & 160V 0.5mTorr & 160V 0.01mTorr & 160V Vacuum \\ \hline
       $\Gamma_e$(m$^{-2}$s$^{-1}$)     & 1.32518E19  & 3.91666E19 & 1.2569E19 & 1.52353E19  \\ \hline
       $\overline{\mathscr{E}}_{e}$(eV) & 8.35 & 75.34  &  82.88  &  83.36 \\ \hline
       $\Gamma_{SEE}$(m$^{-2}$s$^{-1}$) & 4.89249E18 & 3.89133E19 &  1.2568E19 & 1.52353E19\\ \hline
       $\overline{\mathscr{E}}_{SEE}$(eV)& 18.86 & 32.72 &  33.90  & 33.97\\ \hline
       $\Gamma_i$(m$^{-2}$s$^{-1}$)      & 8.35317E18 & 2.53852E17  &  1.9604E15 & 0\\  \hline
       $ \overline{\mathscr{E}}_{i}$(eV) & 64.12 & 11.95 &  0.12 & 0\\ \hline
      \hline
     \end{tabular}
     \label{ElectrodeParameterTable}
    \end{table}
    
  \subsection{GD — glow discharge}
      GD is the traditional sustainable discharge mode of CCP, lying in the inner region of "U" shape of the Pachen curve. 
      A higher ionization rate leads to a high plasma density and causes the obvious sheath-bulk-sheath structure. 
      The difference in electron and ion density caused a strong electric field to be created in the sheath, resulting in a high bulk potential.
      Most of the electrons are confined to the bulk region by this field structure, and the ions in the sheaths are accelerated to bombard the electrodes with higher adjustable energy and flux. 
      Because of this long-term sustainability, high density, and high boundary ion energy, CCP is of great significance in dielectric etching and surface material modification.
      It should be noticed that electron-induced SEE significantly expanded the low-pressure boundary of the "U" shape breakdown curve \cite{smith2003breakdown,radmilovic2005modeling}, shown in figure \ref{BreakdownCondition}, in which the region of the green circle is expanded from the inside of the red line to the right side by a large area.

   \subsection{AM — abnormal multipactor}
       AM is an abnormal sustainable multipactor discharge mode, located in a narrow zone (0.05$\sim$0.5 mTorr, 120$\sim$180 V).
       Because of the low background pressure, the ionization rate is much lower than that of GD. At the same time, low pressure and an rf field result in fast particle mobility, so it is difficult for the particle gathering to increase its density further and form the sheath-bulk-sheath structure like GD. 
       Therefore, compared to GD, the electron densities of AM and NM are much lower than GD. In fact, AM and NM are not even plasma, for the electron density is so low that the Debye length is much larger than the discharge gap. Therefore, the sheath is not formed. However, the distributions of those multipactors are still worthy of attention, and their appearance in microwave devices might be harmful to microwave devices and their performance.
       
       In AM, the source of electrons mainly comes from SEE, resulting in a higher electron density near the electrode and lower in the central, shown in the solid red line of Figure \ref{StableNeiEeiPhi}(a). The lower ionization rate makes it difficult for the electron density to increase further, so the equilibrium is reached quickly before the formation of plasma.
       Therefore, the electron density in AM is almost only 1\% of GD. For a high mass, most ions are trapped in the discharge gap, resulting in a net positive charge that causes the positive potential to be trapped in the gap. 
       Note low electron density is insufficient to shield the applied field. So the shape of the AM potential curve does not have a plateau-like distribution of the GD. 
       This field structure can also accelerate ions to the electrode in a nearly collisionless way, shown in the olive dashed line in figure \ref{StableNeiEeiPhi}(b).
       For a much lower density, the ion flux at the electrode of AM is only 3\% of GD, and the mean ion energy bombarding the electrode is only 11 eV (less than 1/5 of the first type). Therefore, for much lower ion flux, AM cannot be used in etching or deposition like the discharge of GD. 
       Since the applied rf electric field is much higher than the self-generated field, most of the electrons can be accelerated to higher energy and bombarded electrodes to generate new secondary electrons with high coefficients. Therefore, electrons inside the discharge gap can be accelerated to more than 50 eV, as shown in the red line of Figure \ref{StableNeiEeiPhi} (b). Also, the electrons with higher energy are more likely to disappear at the electrode or induce the emission of secondary electrons with lower energy, causing a high core and a low boundary electron energy distribution.

\begin{figure}[ht]
       \centering
         \includegraphics[width=0.5\textwidth]{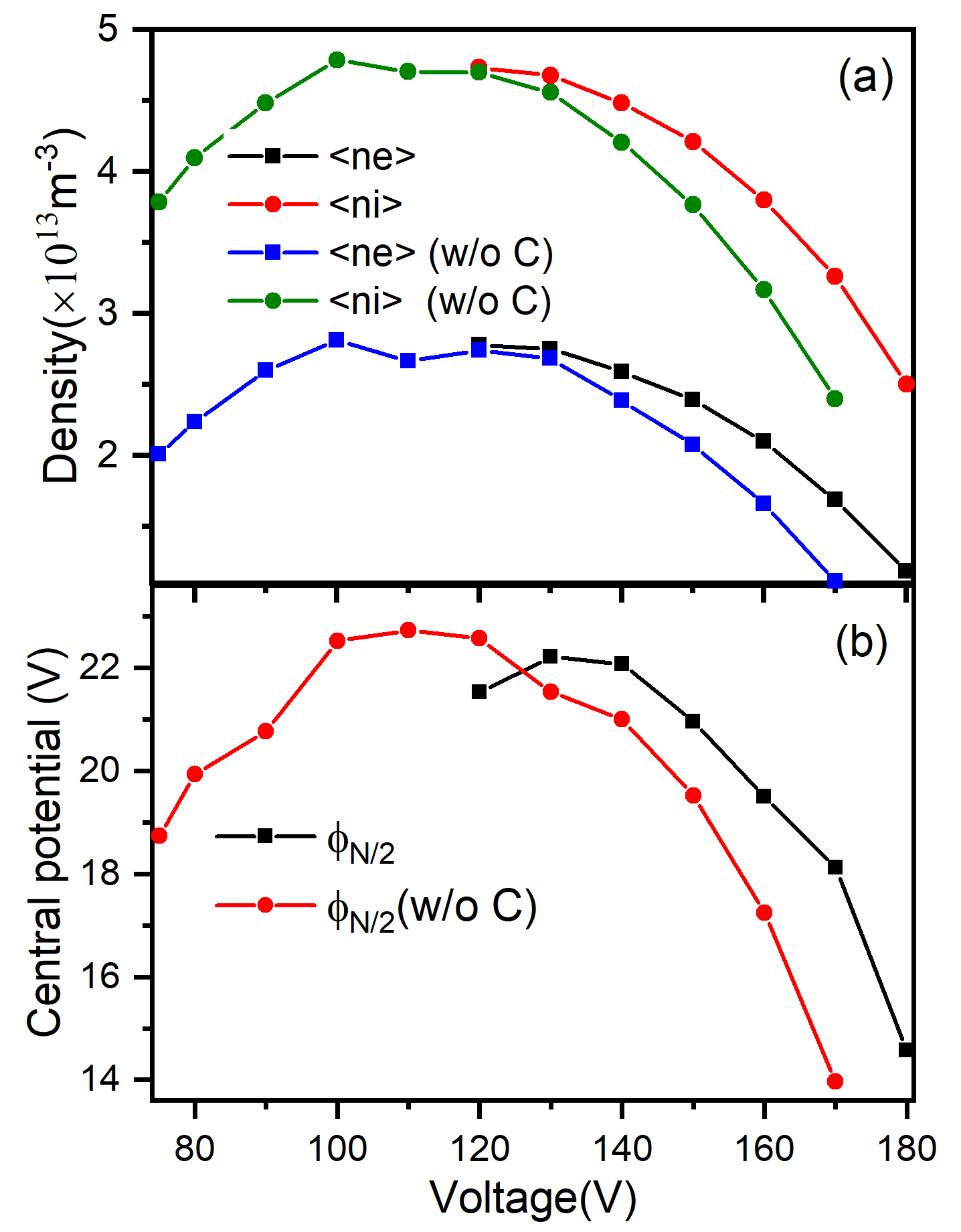}
         \caption{Time and space averaged particle density (a) and time-averaged central potential (b) of AM under different rf voltage (0.5 mTorr), "w/o C" means the results without the 200 pF blocking capacitor}
      \label{AMDiffVoltage}
    \end{figure}

       Effects of rf-voltage on AM in the pressure of 0.5 mTorr are shown in figure \ref{AMDiffVoltage}. Without the blocking capacitor, the discharge parameter range of AM can be expanded from 120$\sim$180 V to 75$\sim$170 V.
       At 0.5 mTorr of AM, electron multiplication effect is maximized by rf power of 100$\sim$120 V, which is consistent with previous experiment work \cite{hohn1997transition}. At the voltage of 100$\sim$120 V, the central voltage reaches its maximum value.
       It should be noticed that the existence of 200 pF blocking capacitor will vanish the left growing curve. The weak oscillating voltage will charge the blocking capacitor and cause bias voltage, which will turn the AM into BFD, and fail the discharge. This may provide a way to prevent the generation of multiplication discharge to a certain extent. 

   \subsection{NM - abnormal multipactor}
     NM is also a sustainable discharge mode, located in the narrow zone of voltage (130$\sim$ 180 V) and the wide range of pressure ($\sim$0.05 mTorr).
     The source of electrons in this mode almost comes from SEE. The ionization collision is so low that it cannot maintain the positive potential of the discharge gap, but it can neutralize the negative charge inside the discharge gap, resulting in zero volts of the mean potential throughout the discharge gap, as shown in the black line of figure \ref{StableNeiEeiPhi}(c).
     Without the confinement of the weaker positive potential in the discharge gap such as AM, and under such an extremely low ionization collision environment, the electron density of the NM is only 20\% of AM.

    Under the force of zero-mean-potential, electrons can be accelerated to more than 60 eV, which bombards the electrode without any confinement. Therefore, the electron energy in the discharge core is a little higher than the electron energy in AM. However, 60MHz rf has almost no effect on ions, so the ions move almost freely in the discharge gap with low energy, shown in the magenta dashed line of figure \ref{StableNeiEeiPhi}(b).
    
\begin{figure}[ht]
       \centering
         \includegraphics[width=0.5\textwidth]{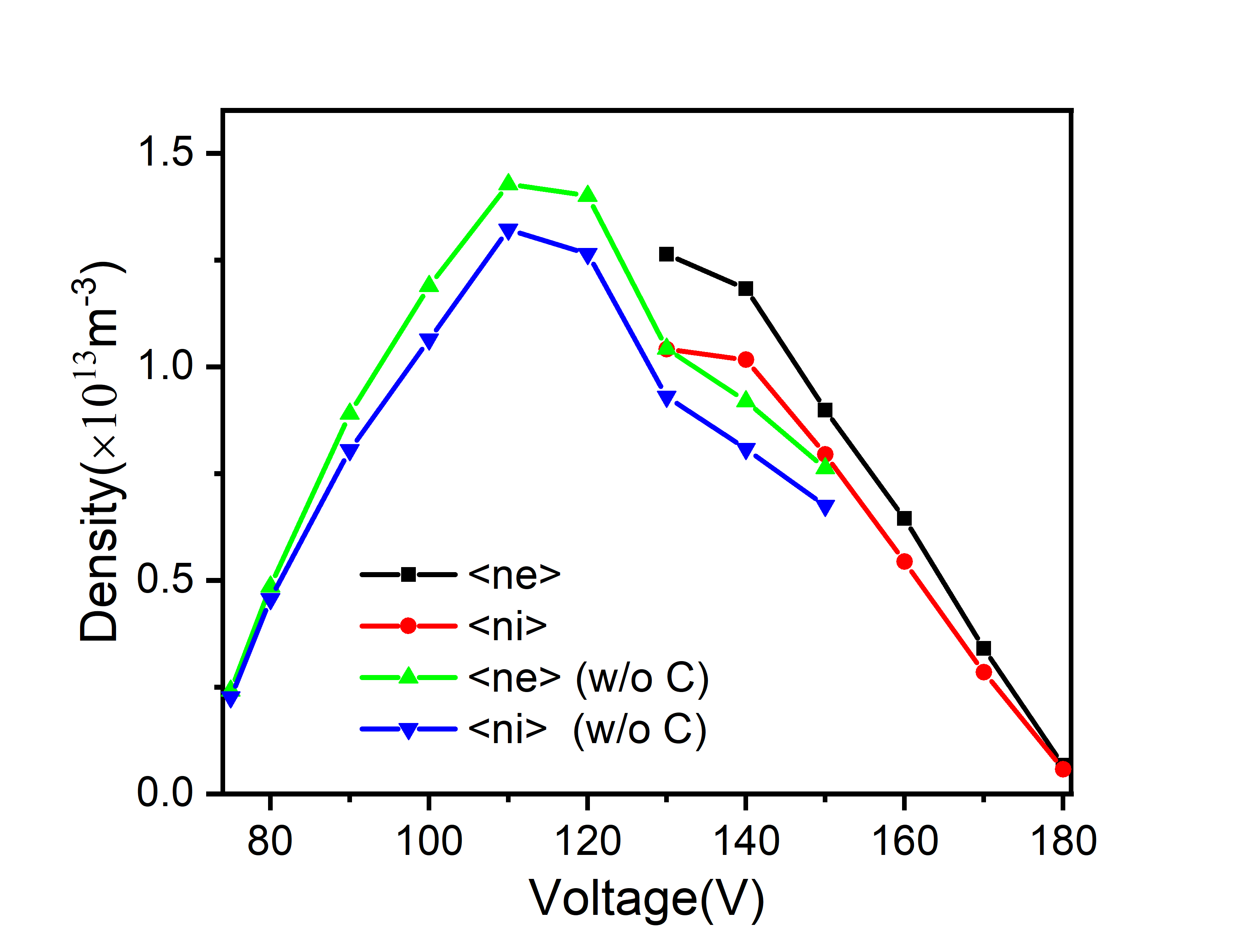}
         \caption{Time and space averaged particle density of NM under different rf voltage, "w/o C" means the results without the 200 pF blocking capacitor}
      \label{NMDiffVoltage}
    \end{figure}

    The effect of voltage on the discharge of NM is similar to AM, as shown in figure \ref{AMDiffVoltage}(a) and figure \ref{NMDiffVoltage}. Without the blocking capacitor, the particle density grows with the increases of voltage in 75$\sim$110 V, and declines in 110$\sim$150 V. However, without the positive central potential, the linear characteristics of VM are more obvious than AM, which comes from the force of the applied external field at low ion density. 
    It should be noticed that electron density is close but slightly higher than ion density. Such low net negative charge density almost has no effect potential.
    
    In the pressure of 0.001$\sim$0.05 mTorr, the existence of blocking capacitor can also vanish the rising part of the density-voltage curve, as shown in figure \ref{NMDiffVoltage}. 
    The unsustainable discharge in 80$\sim$130 V caused by blocking capacitor turns the NM to BFD just like the effect of blocking capacitor on AM. Therefore, in the case of 60 MHz, 200 pF DC blocking capacitor can greatly narrow the zone of multipactor discharge parameters.

 \subsection{VM — vacuum multipactor}
    When the pressure of the discharge chamber drops to 0 (achieved by closing the Monte Carlo module), it is in absolute vacuum condition. Under the force of an oscillating electric field, the discharge can be sustained by a high boundary of the SEE coefficient. Thus, the electron density maintains a higher value near the boundary.
    The kind of discharge in this case has an electron density distribution similar to that of the AM and the NM. However, zero pressure means without the source of positive ions, which caused a weak negative potential in the discharge gap. The negative potential of the week will push the electron to both sides of the electrodes, resulting in a lower electron density inside the discharge gap.

\section{Discussion—formation of sustainable discharges} \label{sec4}
   \subsection{General consideration} 
    Usually, multipactor can hardly be found in the study of rf-driven CCP. In the experiment, due to the extremely low electron density of AM and NM and the relatively thin background gas, it is difficult to observe the luminescence phenomenon during those discharges, and it is likely to be treated as unable to discharge in the experiment. This may also be the main reason why multipactor has been neglected in the experiment of rf-driven CCP. 
    It can also hardly be found in the numerical simulation of rf-driven CCP. On the one hand, in the steady-state simulation, to reach the last stable discharge state as fast as possible, the higher plasma density is chosen as the initial density condition, which will form the sheath in the initial time, and will make it easy to ignore AM and NM in the discharge of CCP. Only evolutionary simulation can distinguish these three types of sustainable discharge under different discharge conditions.
    On the other hand, ESEE, especially the high emission coefficient ESEE used for SiO$_2$ or metal oxides is seldom considered, which directly makes the multipactor difficult to find in simulating rf-driven CCP. 
    In addition, in the study of the Paschen curve, because of the existence of the RFD region, the GD and its left region are likely to be considered as impenetrable regions. Therefore, discharge signatures such as AM and NM appear only through evolutionary simulations with very low initial densities and only when conditions such as higher frequency ranges, specific voltages, specific geometric spacings, and very low gas pressures are met. 
    With the help of the improved evolutionary simulation code, these three types of discharge will be further discussed and compared to reveal the formation of these discharge modes.
     
\begin{figure}[ht]
       \centering
         \includegraphics[width=0.5\textwidth]{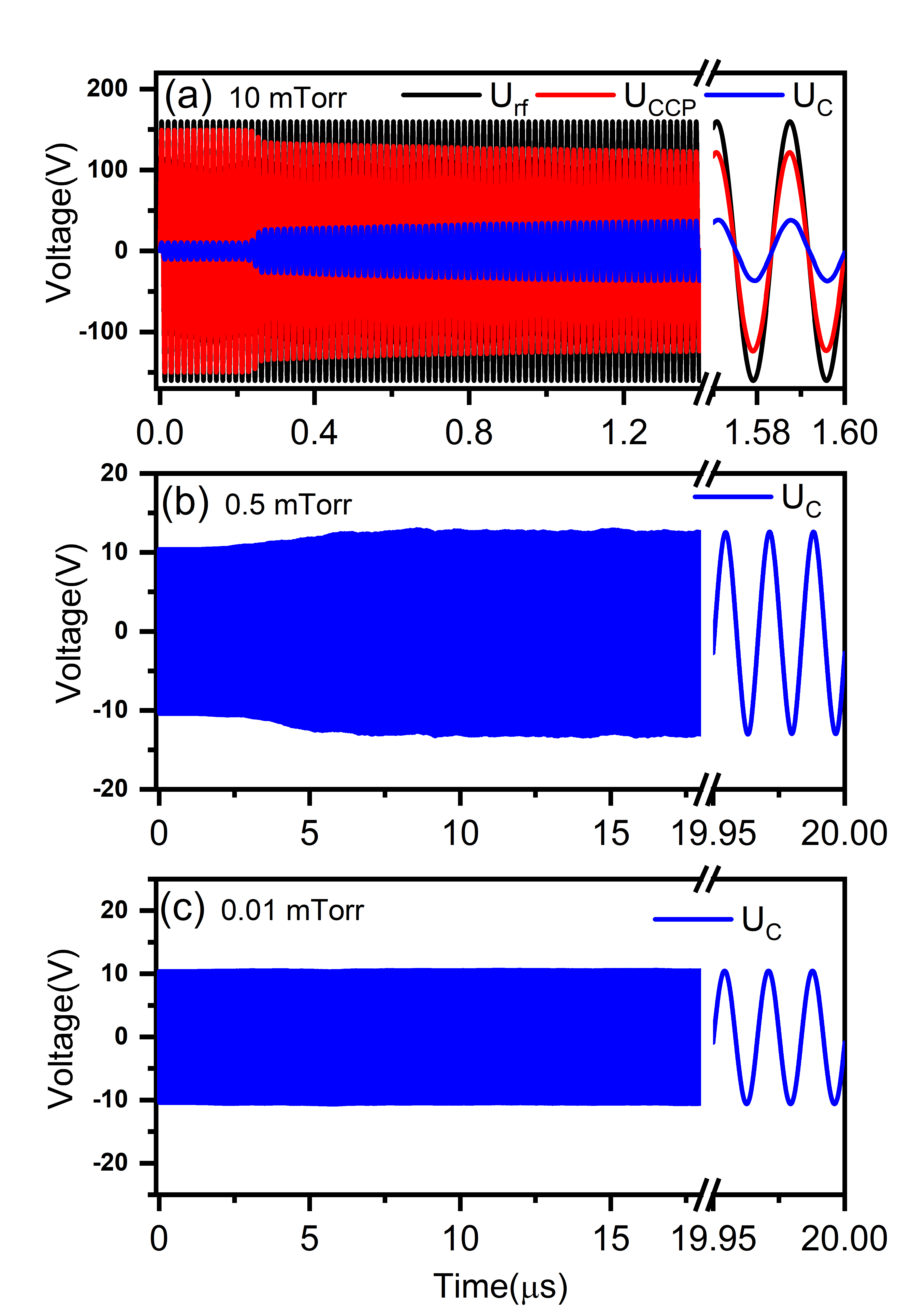}
         \caption{Evolution of the circuit signal, (a) GD (10mTorr, 160V), (b) AM (0.5 mTorr, 160V), (c) NM (0.01mTorr, 160V)}
      \label{StableECUsUccpUc}
    \end{figure}

   The electric voltage waveform of the three sustainable discharge are shown in figure \ref{StableECUsUccpUc}. The formation of sheath-bulk-sheath directly declines the voltage of the electrode, proving the effective electrical character has been changed to a great extent during the rf breakdown, which has been analyzed in Ref.\cite{wu2021electrical,wu2022effects}. For AM, the formation of positive potential has little effect on the signal of the outside circuit. Only 2$\sim$4 V voltage amplitude incline for blocking capacitor during the forming of AM. For VM, for low electron density and near zero net charge distribution, the self-generated electric field can almost neglected. Therefore, the discharge chamber of VM can almost be treated as a vacuum parallel plate capacitor. So the signal has no change during the formation of NM.

\begin{figure}[ht]
       \centering
         \includegraphics[width=\textwidth]{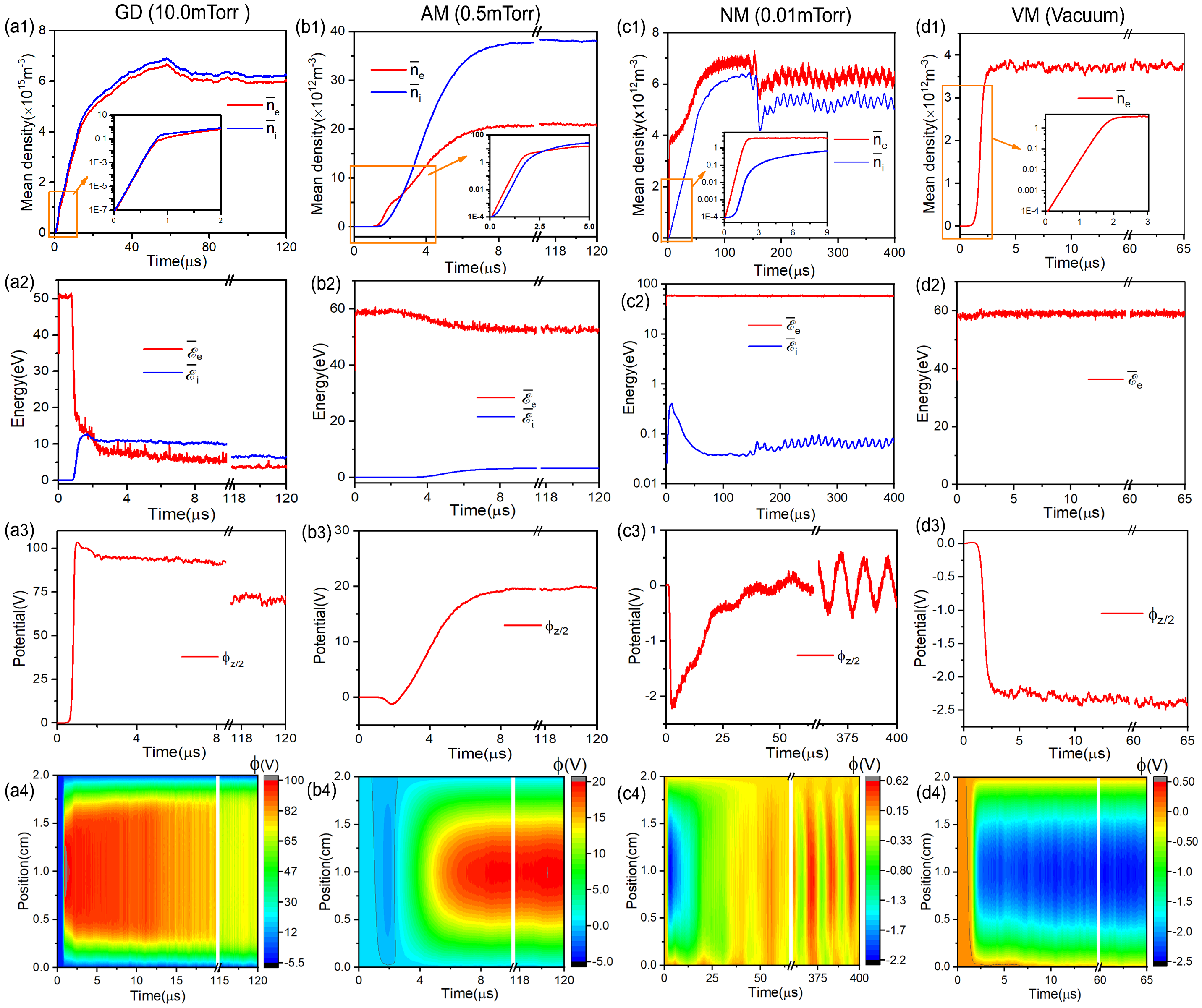}
         \caption{The evolution of particle mean density (1), energy(2), and potential (3,4) in different discharge modes. (a) GD, (b) AM, (c) NM, and (d) is the VM}
      \label{NeNiTeTiComparing}
    \end{figure}

   The external circuit electrical signal can reflect the formation process of those discharges to a certain extent. However, to study the reasons for their formation, it is necessary to discuss the evolution of the parameters inside the discharge chamber.
   The evolution of the average electron density, electron energy, core potential, and potential spatial distribution of various discharges is shown in figure \ref{NeNiTeTiComparing}. With the help of figure \ref{NeNiTeTiComparing} and the time-space evolution of the density and temperature of various discharges, as well as the curves of the particle number balance and power evolution, the discharges of GD, AM, NM, and VM will be thoroughly analyzed in this section.
    
 \subsection{formation of GD} 
\begin{figure}[ht]
       \centering
         \includegraphics[width=\textwidth]{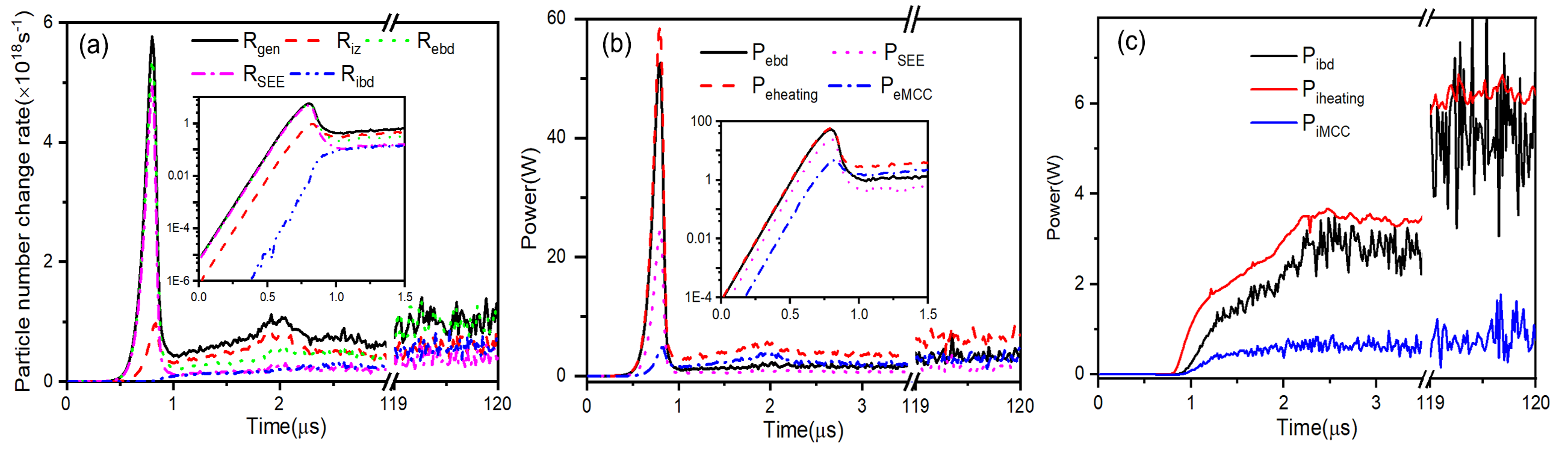}
         \caption{Temporal evolution of particle change rate and Power}
      \label{10.0PPBalance}
    \end{figure}

   In the initial time, for low electron density, the electric field can penetrate the entire discharge gap directly and accelerate the electrons. Therefore, most electrons maintain high energy (about 80eV, shown in figure \ref{NeNiTeTiComparing}(a2)). The energy gained from electrons will collide and ionize the background argon to generate a new electron-ion pair. At the same time, plenty of electrons will bombard the electrode with high energy (TABLE.\ref{ElectrodeParameterTable}) to generate SEE with a high emission coefficient\cite{vender1996simulations,wu2021electrical}. 
    \begin{figure}[ht]
       \centering
         \includegraphics[width=0.6\textwidth]{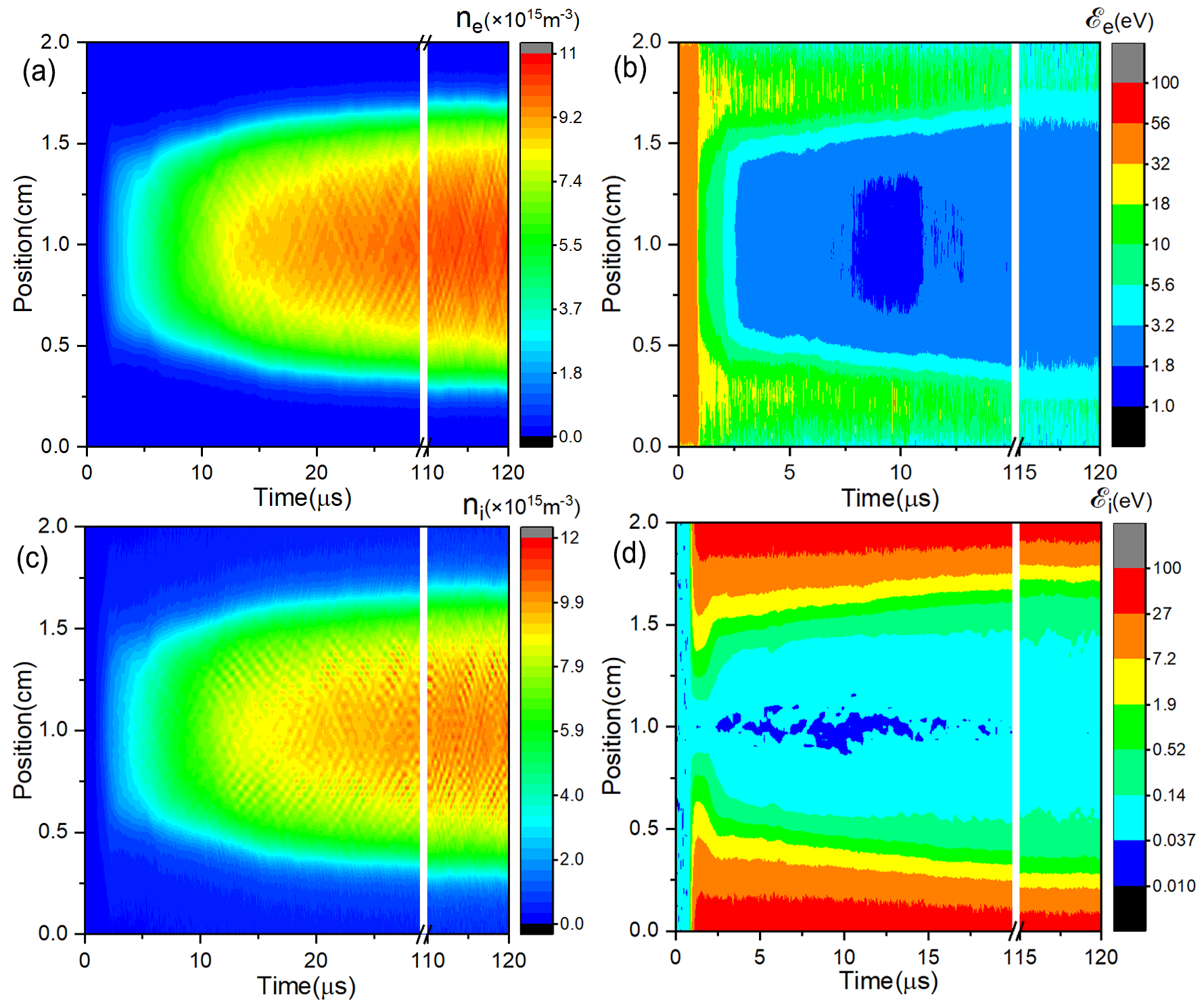}
         \caption{Spatio-Temporal evolution of electron density (a), electron energy (b), ion density (c), ion energy(d)}
      \label{10.0STevolution}
    \end{figure}
    A high ionization rate and an ESEE rate exponentially increase the electron density.
    Usually, on the right side of the GD region, for higher pressure, ionization dominated the discharge, which is shown in the breakdown study of Ref.\cite{wu2021electrical}; On the left side of the GD zone, however, ESEE occupies an indispensable position of the lower pressure, as shown in Figure \ref{10.0PPBalance}(a). When $n_e$ is high enough, electrons near the electrodes escape quickly and the sheath forms quickly, which maximizes the particle generation rate and electron absorption power. The gas, at this time, can be treated for breakdown.

    With the stabilization of two sheaths, electrons quickly cooled, and the growth rate of the plasma density slowed down and stabilizes. The presence of the sheath, in turn, limits the loss of electrons until the bipolar diffusion reaches equilibrium and a more obvious sheath-bulk-sheath structure is formed. Most electrons are confined in the bulk region, and ions are gradually accelerated by the sheath electric field to the electrodes. 
    The particle and energy balance has been discussed in detail in Ref.\cite{wu2021electrical}, which is quite similar to figure \ref{10.0PPBalance}. However, it should be noted that the background gas pressure is much lower, and a higher SEE coefficient is considered in this work, even if the stable discharge is reached, the secondary electron in particle and power balance still occupies a high proportion, as shown in the figure \ref{10.0PPBalance}(b,c). 

    The pressure results shown in Figure \ref{10.0PPBalance} and Figure \ref{10.0STevolution} is lower than 10 mTorr, which is much lower than in Ref.\cite{wu2021electrical}, and the discharge gap is also narrower, which makes the particle boundary force more obvious. Thus, the boundary emission opp particle occupies a non-negligible proportion. When the gas pressure grows, the proportion of an electron source will gradually be dominated by ionization.

\subsection{formation of AM} 
   When the background gas pressure is as low as 0.5 mTorr, in the initial time, the electric field can still penetrate the entire discharge gap. Under the force of this penetrable electric field, most electrons will gain energy and bombard the electrode. 
   A large amount of secondary electron emission from the surface dominates the initial discharge. In a short period of time, the weak ionization rate can be almost ignored in such low pressure. The penetrable oscillating electric field accelerates electrons directly, maintaining a higher energy state of electrons, shown in figure \ref{NeNiTeTiComparing}(b2). 
   The high secondary electron emission rate(>1) in both electrode lead to the exponential growth of the electron density before 2.5 µs, shown in figure \ref{NeNiTeTiComparing}(b1). 
   
   The accumulation of electrons results in a weaker negative potential in the discharge gap, shown in Figure \ref{NeNiTeTiComparing}(b3,b4), which will push the electrons to the electrode, increasing the electron boundary loss rate growing rapidly.
   The mass of the electron is so light that its drift and diffusion current can reach the equilibrium first before 2 µs.
   This first equilibrium is sustained by the electron emission and absorption in the boundary, and the force of oscillating electric field.
   The electron energy in the discharge gap is as high as 60 eV, which is much higher than the threshold of ionization collision (15.8 eV).  
   
\begin{figure}[ht]
       \centering
         \includegraphics[width=\textwidth]{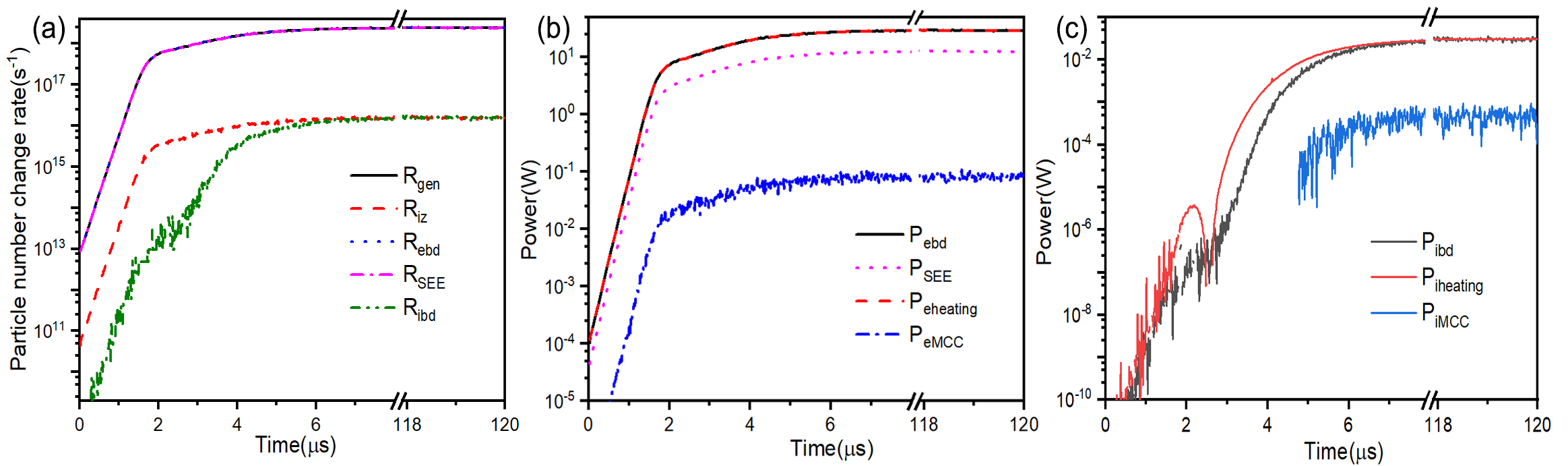}
         \caption{Temporal evolution of particle change rate and Power}
      \label{0.5PPBalance}
    \end{figure}

   After 2 µs, both the growth rate of electron density and the power deviate from the original exponential trend of the avalanche and slowed significantly, shown in figure \ref{NeNiTeTiComparing}(b1) and figure \ref{0.5PPBalance}(a,b).
   Even if the gas pressure is 0.5 mTorr, many ionization collisions still occur inside the gap, which generates plenty of ions. For a high mass, the rf field has almost no effect on the ions. Under the force of the initial negative potential region around 2 µs, ions are trapped inside the core and the ion density slowly increases. The accumulating Ar$^+$ ions quickly neutralize the negative potential. It will take a few microseconds to move to the boundary in the zero-potential region, which is greater than the relaxation time of the electron-ionization collision. This will result in the accumulation of a small number of ionized ions in the discharge gap, leading to a higher positive potential as shown in \ref{NeNiTeTiComparing}(b3,b4). Gradually, the ion density is greater than the electron density, shown in figure \ref{NeNiTeTiComparing}(b1). 

      \begin{figure}[ht]
       \centering
         \includegraphics[width=0.6\textwidth]{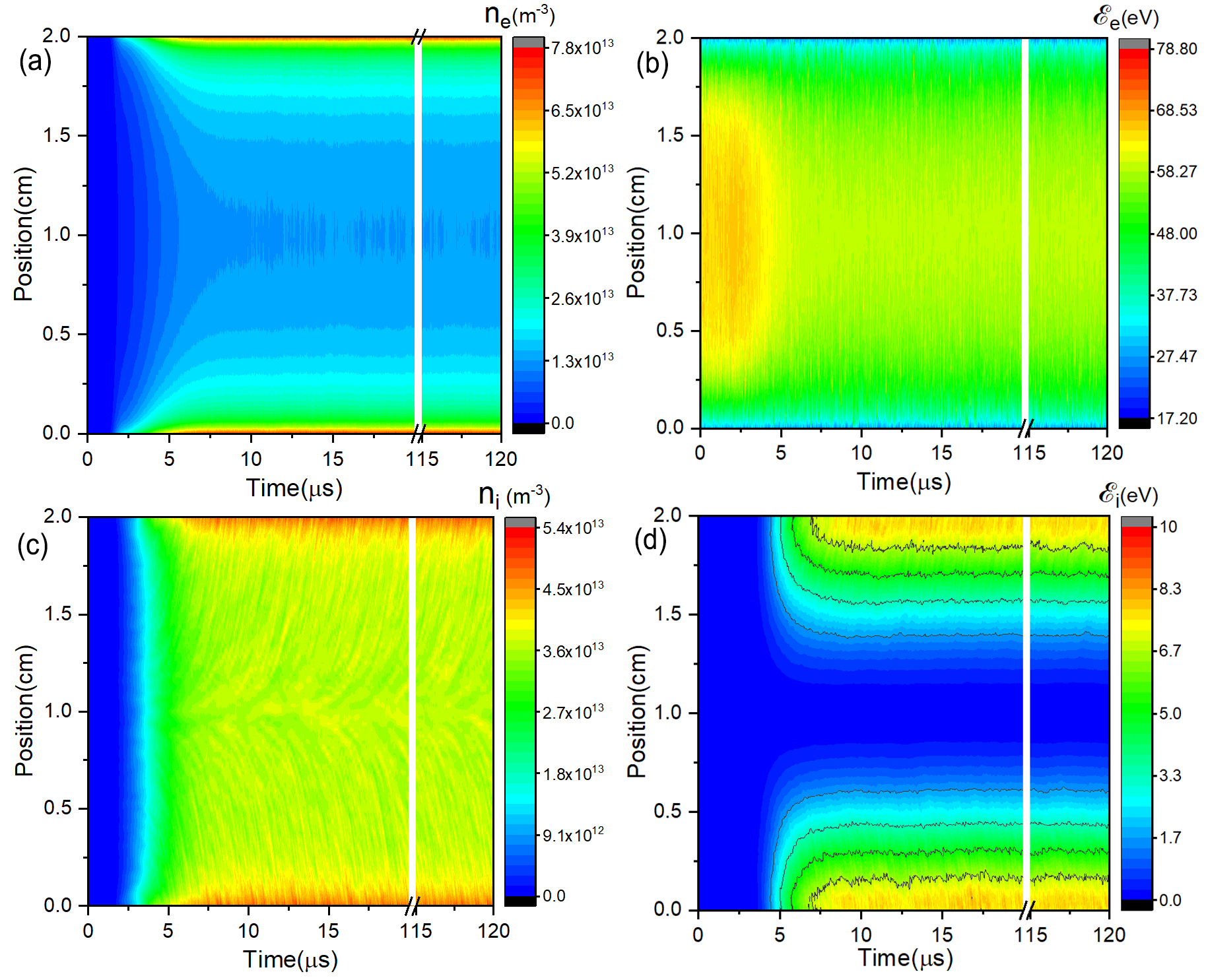}
         \caption{Spatio-Temporal evolution of electron density (a), electron energy (b), ion density (c), ion energy(d)}
      \label{0.5STevolution}
    \end{figure}

   A large amount of positive ions accumulated in the gap will cause the potential to grow, which in turn traps part of the electrons to the core and pushes the ions to the electrode. Therefore, under the force of the growth of positive potential, the loss of the boundary, the absorption power, and the loss of the boundary power of the ions gradually increase, shown in Figure \ref{0.5PPBalance} (a,c).
   Finally, when the growth ion loss rate at the boundary and ionization rate inside the discharge gap equal, the particle number and power absorption of the ion also reach their balance.
   The positive potential caused by ion growth, ionization collisions, and the positive potential caused by the second growth in electron density. The electron density has increased several times and reached the last equilibrium again. This time the equilibrium is maintained by stronger surface emission, weaker gap ionization, stronger oscillating electric field, and weaker self-generated field.

\subsection{formation of NM}

\begin{figure}[ht]
       \centering
         \includegraphics[width=0.6\textwidth]{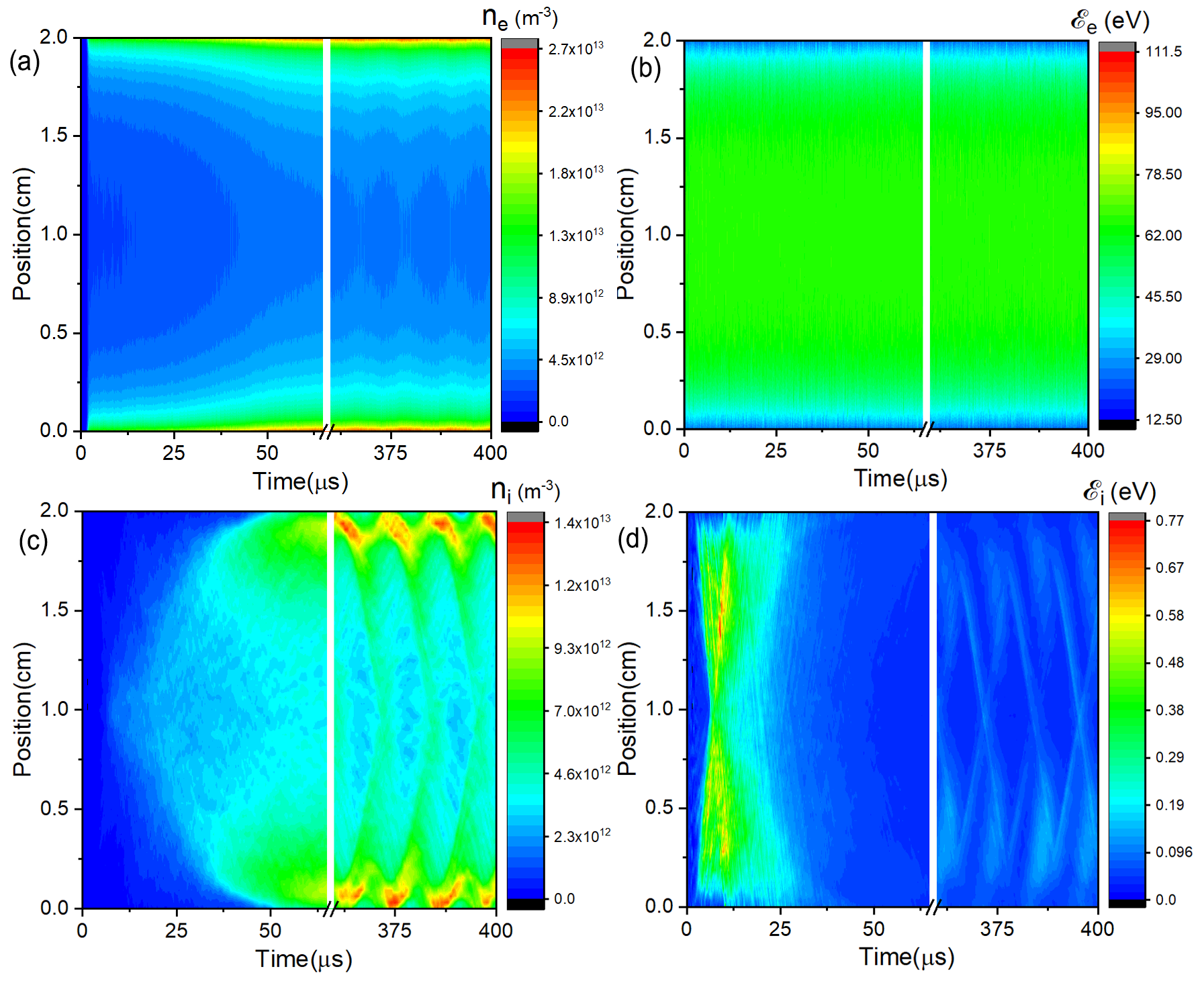}
         \caption{Spatio-Temporal evolution of electron density (a), electron energy (b), ion density (c), ion energy(d)}
      \label{0.01STevolution}
    \end{figure}

   When extremely low pressures are chosen, at the initial moment of the NM, electron dynamics is almost the same as AM, the electrons and related power of it reach the first equilibrium within 2µs.
   However, the gas pressure is so low that the extremely low ionization rate can not generate enough ions to neutralize the negative potential as fast as AM. However, there will still be a small number of positive ions produced, which are confined in the discharge gap for a long time. In this phase, those ions trapped in the weak negative potential region will also oscillate under the force of the rf electric field and gain energy, resulting in the higher ion energy before 25µs(shown in figure \ref{NeNiTeTiComparing}(c2) and figure \ref{0.01STevolution}(d)). 
   
   Gradually, the new ions will also be trapped and accumulated in the core, resulting in the ion density growing up to the electron density. This process will take longer. At about 50 µs in 0.01 mTorr, the negative potential is almost neutralized and the discharge enters the second equilibrium.
   This process also increases the electron density from the original $4\times10^{12}$ m$^{-3}$ of the first equilibrium to $7\times10^{12}$ m$^{-3}$ of the second equilibrium, as shown in figure \ref{NeNiTeTiComparing}(c1).
   
\begin{figure}[ht]
       \centering
         \includegraphics[width=\textwidth]{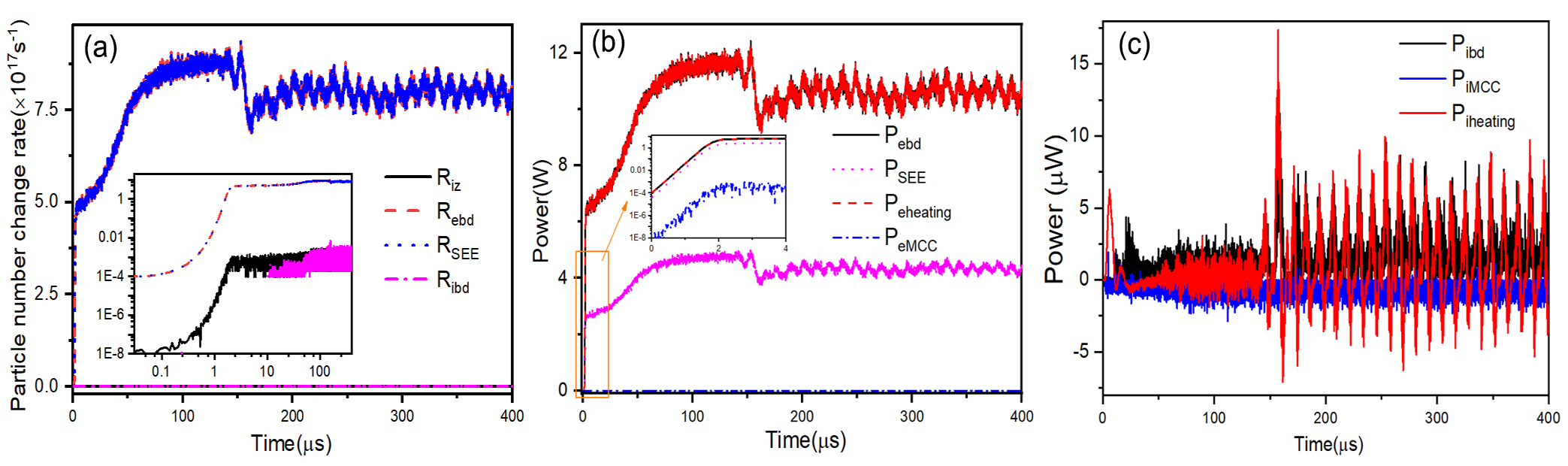}
         \caption{Temporal evolution of particle change rate and Power}
      \label{0.01PPBalance}
    \end{figure}

   Because the negative potential is neutralized, ions can move freely and gradually diffuse to the two electrode. In the zero-potential region, some ions gradually diffuse to the boundary, causing the ion loss rate at the boundary to gradually increase and gradually equal to the ionization rate, and the number of ions gradually reaches equilibrium.
   At about 100 µs, the ion generating rate from the ionization collision($R_{iz}$) is gradually equal to the boundary loss rate($R_{ibd}$), shown in figure \ref{0.01PPBalance}(a). The power also shows the trend of equilibrium and stability. The energy of the electron is provided by the electric field and the SEE (almost half of each term), and most of the power is lost at the boundary. 
   
   The discharge system seems to have reached their respective equilibrium at about 100µs and looks like will be sustained for a long time within 100$\sim$150 µs
   However, the equilibrium built under zero potential is fragile. The randomness of ion boundary loss, secondary electron emission, and the randomness of MCC collisions can disturb the equilibrium, which will cause space-charge accumulation, boundary capacitor charging, etc., causing resonance of ions and fields.
   Fortunately, a higher electric field disturbance caused by the gathering of ions will suppress further enhancement of the unstable feature, causing the ion density to be in a dynamic oscillation distribution. The frequency of this oscillation is lower than 0.1 MHz.

 \subsection{formation of VM} 

   When the pressures drop to zero (absolute vacuum), there will be no generation of ions. In this simulation, the absolute vacuum environment can be achieved by turning off the MCC module. In the discharge where only electrons exist, the electron number and its related power will reach equilibrium quickly within 2.0 µs, which is similar to the initial stage of AM and NM. The difference is that there will be no generation of ions, and this equilibrium will not change thereafter.
  
   However, it should be noticed that it is impossible to reach the absolute vacuum in the experiment. When the pressure is less than 0.05mTorr, in the ESEE model with a high emission coefficient, NM will be the most common discharge under extremely low background pressure. So the VM is just the result of theory.
    
\section{Conclusion} \label{sec5}
   In this work, we used a one-dimensional implicit PIC/MCC model to study the capacitive-like sustainable discharges driven by rf power in the background gas of pure argon. The electron-induced secondary emission model with high coefficients that is suitable for SiO$_2$ electrode is taken into account. 
   
  After scanning the wide range of voltage-pressure zone, we found that on the left side of Paschen's curve, with the decline of gas pressure, the discharge will be transformed from glow discharge to abnormal multipactor with positive potential inside the gap, and transformed to normal multipactor with zero potential when lower than 0.1 mTorr. A higher ionization rate resulted in an abnormal one, which caused a positive potential of several volts to tens of volts in the core. Weak ionization results in zero potential. The electron sources of these two-type multipactor mainly come from secondary electron emission.
  These new discharges have higher electron energy and electron flux at the boundary and are mainly sustained by higher electrode-induced SEE coefficient and rf power with high frequency.

  The electron multiplication phenomenon is most obvious around 120V, in which electron density reaches the maximum. The existence of 200 pF blocking capacitor will create tens of bias volts in lower rf voltage region and fail the discharge.

  We then discussed the formation processes of three different discharge modes in detail under the evolutionary simulating methods.
  The formation of the three discharge modes is dominated by electron dynamics. Without the background gas (vacuum), pure electron multiplication process will result in negative potential distribution. With the background gas, different growth rate of ion density caused by ionization gradually differentiates them: a high ionization in glow discharge creates strong potential to gather electrons until sheath-bulk-sheath structure formed and stabilized; weak ionization in abnormal multipactor can only create tens of volts of the potential; extremely low ionization rate in normal multipactor can only neutralize the negative potential caused by initial electron multiplication of the gap. Zero potential in the normal multipactor will cause low-frequency oscillation of ion density distribution.
  
  These new discharge modes might broaden the theory of gas discharge and expand the application of rf-driven discharge with SiO$_2$ or metal oxide film (such as aluminum oxide) applied electrodes.


\section*{Acknowledgments}
This work was supported by the National Natural Science Foundation of China (12275095, 11975174 and 12011530142), and the Fundamental Research Funds for Central Universities (WUT: 2020IB023), and the Hubei Provincial Natural Science Foundation of China (2023AFB488),
and Hubei University of Science and Technology Doctoral Startup Foundation (BK202401).

\section*{Data availability statement}
The data that support the findings of this study are available from the corresponding author upon reasonable request.

\section*{ORCIDs}
Hao Wu https://orcid.org/0000-0003-1074-6853\\
Ran An https://orcid.org/0000-0003-2692-401X\\
Wei Jiang https://orcid.org/0000-0002-9394-585X\\
Ya Zhang https://orcid.org/0000-0003-0473-467X\\

\nolinenumbers

\bibliography{library}

\providecommand{\noopsort}[1]{}\providecommand{\singleletter}[1]{#1}%
\begin{thebibliography}{10}

\bibitem{lieberman_principles_2005}
Michael~A. Lieberman and Allan~J. Lichtenberg.
\newblock {\em Principles of Plasma Discharges and Materials Processing:
  Lieberman/Plasma 2e}.
\newblock John Wiley \& Sons, Inc., 2005.

\bibitem{chabert2011physics}
Pascal Chabert and Nicholas Braithwaite.
\newblock {\em Physics of radio-frequency plasmas}.
\newblock Cambridge University Press, 2011.

\bibitem{chen2023note}
Lei Chen, Hao Wu, Zili Chen, Yu~Wang, Lin Yi, Wei Jiang, and Ya~Zhang.
\newblock Note on particle balance in particle-in-cell/monte carlo model and
  its implications on the steady-state simulation.
\newblock {\em Plasma Sources Science and Technology}, 32(3):034001, 2023.

\bibitem{wu2022note}
Hao Wu, Zhaoyu Chen, Lin Yi, Wei Jiang, and Ya~Zhang.
\newblock Note on the energy transport in capacitively coupled plasmas.
\newblock {\em Plasma Sources Science and Technology}, 31(4):047001, 2022.

\bibitem{phelps1999cold}
AV~Phelps and Z~Lj Petrovic.
\newblock Cold-cathode discharges and breakdown in argon: surface and gas phase
  production of secondary electrons.
\newblock {\em Plasma Sources Science and Technology}, 8(3):R21, 1999.

\bibitem{verboncoeur2005particle}
John~P Verboncoeur.
\newblock Particle simulation of plasmas: review and advances.
\newblock {\em Plasma Physics and Controlled Fusion}, 47(5A):A231, 2005.

\bibitem{daksha2019material}
Manaswi Daksha, Aranka Derzsi, Z~Mujahid, David Schulenberg, Birk Berger,
  Zolt{\'a}n Donk{\'o}, and Julian Schulze.
\newblock Material dependent modeling of secondary electron emission
  coefficients and its effects on pic/mcc simulation results of capacitive rf
  plasmas.
\newblock {\em Plasma Sources Science and Technology}, 28(3):034002, 2019.

\bibitem{lafleur2013secondary}
Trevor Lafleur, Pascal Chabert, and Jean-Paul Booth.
\newblock Secondary electron induced asymmetry in capacitively coupled plasmas.
\newblock {\em Journal of Physics D: Applied Physics}, 46(13):135201, 2013.

\bibitem{sydorenko2006kinetic}
D~Sydorenko, A~Smolyakov, I~Kaganovich, and Y~Raitses.
\newblock Kinetic simulation of secondary electron emission effects in hall
  thrusters.
\newblock {\em Physics of Plasmas}, 13(1):014501, 2006.

\bibitem{kaganovich2007kinetic}
ID~Kaganovich, Yevgeny Raitses, Dmytro Sydorenko, and Andrei Smolyakov.
\newblock Kinetic effects in a hall thruster discharge.
\newblock {\em Physics of Plasmas}, 14(5):057104, 2007.

\bibitem{kishek1998multipactor}
Rami~A Kishek, YY~Lau, LK~Ang, Agust Valfells, and Ronald~M Gilgenbach.
\newblock Multipactor discharge on metals and dielectrics: Historical review
  and recent theories.
\newblock {\em Physics of Plasmas}, 5(5):2120--2126, 1998.

\bibitem{horvath2017role}
B~Horv{\'a}th, M~Daksha, I~Korolov, A~Derzsi, and J~Schulze.
\newblock The role of electron induced secondary electron emission from sio2
  surfaces in capacitively coupled radio frequency plasmas operated at low
  pressures.
\newblock {\em PLASMA SOURCES SCIENCE \& TECHNOLOGY}, 26(12), 2017.

\bibitem{horvath2018effect}
Benedek Horv{\'a}th, Julian Schulze, Zolt{\'a}n Donk{\'o}, and Aranka Derzsi.
\newblock The effect of electron induced secondary electrons on the
  characteristics of low-pressure capacitively coupled radio frequency plasmas.
\newblock {\em Journal of Physics D: Applied Physics}, 51(35):355204, 2018.

\bibitem{vender1996simulations}
D~Vender, HB~Smith, and RW~Boswell.
\newblock Simulations of multipactor-assisted breakdown in radio frequency
  plasmas.
\newblock {\em Journal of applied physics}, 80(8):4292--4298, 1996.

\bibitem{smith2003breakdown}
HB~Smith, Christine Charles, and RW~Boswell.
\newblock Breakdown behavior in radio-frequency argon discharges.
\newblock {\em Physics of Plasmas}, 10(3):875--881, 2003.

\bibitem{radmilovic2005modeling}
M~Radmilovi{\'c}-Radjenovi{\'c} and JK~Lee.
\newblock Modeling of breakdown behavior in radio-frequency argon discharges
  with improved secondary emission model.
\newblock {\em Physics of plasmas}, 12(6):063501, 2005.

\bibitem{wu2021electrical}
Hao Wu, Youyou Zhou, Jiamao Gao, Yanli Peng, Zhijiang Wang, and Wei Jiang.
\newblock Electrical breakdown in dual-frequency capacitively coupled plasma: A
  collective simulation.
\newblock {\em Plasma Sources Science and Technology}, 2021.

\bibitem{lisovskiy1998rf}
VA~Lisovskiy and VD~Yegorenkov.
\newblock Rf breakdown of low-pressure gas and a novel method for determination
  of electron-drift velocities in gases.
\newblock {\em Journal of Physics D: Applied Physics}, 31(23):3349, 1998.

\bibitem{hohn1997transition}
F~H{\"o}hn, W~Jacob, R~Beckmann, and R~Wilhelm.
\newblock The transition of a multipactor to a low-pressure gas discharge.
\newblock {\em Physics of Plasmas}, 4(4):940--944, 1997.

\bibitem{kim2006transition}
HC~Kim and JP~Verboncoeur.
\newblock Transition of window breakdown from vacuum multipactor discharge to
  rf plasma.
\newblock {\em Physics of plasmas}, 13(12):123506, 2006.

\bibitem{udiljak2003new}
Richard Udiljak, D~Anderson, P~Ingvarson, U~Jordan, U~Jostell, L~Lapierre,
  G~Li, M~Lisak, J~Puech, and J~Sombrin.
\newblock New method for detection of multipaction.
\newblock {\em IEEE transactions on plasma science}, 31(3):396--404, 2003.

\bibitem{na2019analysis}
Dong-Yeop Na and Fernando~L Teixeira.
\newblock Analysis of multipactor effects by a particle-in-cell algorithm
  integrated with secondary electron emission model on irregular grids.
\newblock {\em IEEE Transactions on Plasma Science}, 47(2):1269--1278, 2019.

\bibitem{zhang2019analytical}
Ziyi Zhang, Yanzi Sun, Wanzhao Cui, Hongtai Zhang, Yindong Huang, and Chao
  Chang.
\newblock An analytical model of one-sided multipactor on a dielectric of a
  metal surface for spacecraft application.
\newblock {\em IEEE Transactions on Electron Devices}, 66(11):4921--4927, 2019.

\bibitem{hubble2017multipactor}
Aimee~A Hubble, Vernon~H Chaplin, Kathryn~A Clements, Rostislav Spektor,
  Preston~T Partridge, and Timothy~P Graves.
\newblock Multipactor breakdown threshold reduction due to magnetic confinement
  in parallel fields.
\newblock {\em IEEE Transactions on Plasma Science}, 45(7):1726--1730, 2017.

\bibitem{spektor2018space}
R~Spektor, MS~Feldman, AA~Hubble, and TP~Graves.
\newblock Space charge saturation in multipactor discharges with parallel
  magnetic field.
\newblock {\em Physics of Plasmas}, 25(12):122109, 2018.

\bibitem{feldman2018effects}
Matthew~S Feldman, Aimee~A Hubble, Rostislav Spektor, and Preston~T Partridge.
\newblock Effects of backscattered electrons on multipactor simulations with
  parallel magnetic fields.
\newblock In {\em 2018 IEEE MTT-S International Conference on Numerical
  Electromagnetic and Multiphysics Modeling and Optimization (NEMO)}, pages
  1--3. IEEE, 2018.

\bibitem{iqbal2023two}
Asif Iqbal, De-Qi Wen, John Verboncoeur, and Peng Zhang.
\newblock Two surface multipactor with non-sinusoidal rf fields.
\newblock {\em Journal of Applied Physics}, 134(15), 2023.

\bibitem{iqbal2023recent}
Asif Iqbal, De-Qi Wen, John Verboncoeur, and Peng Zhang.
\newblock Recent advances in multipactor physics and mitigation.
\newblock {\em High Voltage}, 2023.

\bibitem{wen2022higher}
De-Qi Wen, Peng Zhang, Janez Krek, Yangyang Fu, and John~P Verboncoeur.
\newblock Higher harmonics in multipactor induced plasma ionization breakdown
  near a dielectric surface.
\newblock {\em Physical Review Letters}, 129(4):045001, 2022.

\bibitem{guo2019secondary}
Junjiang Guo, Dan Wang, Yantao Xu, Xiangping Zhu, Kaile Wen, Guanghui Miao,
  Weiwei Cao, JinHai Si, Min Lu, and Haitao Guo.
\newblock Secondary electron emission characteristics of al2o3 coatings
  prepared by atomic layer deposition.
\newblock {\em AIP advances}, 9(9):095303, 2019.

\bibitem{wu2022breakdown}
Hao Wu, Zhaoyu Chen, Zhijiang Wang, Bo~Rao, Wei Jiang, and Ya~Zhang.
\newblock On the breakdown process of capacitively coupled plasma in carbon
  tetrafluoride.
\newblock {\em Journal of Physics D: Applied Physics}, 55(25):255203, 2022.

\bibitem{vahedi1993capacitive}
V~Vahedi, G~DiPeso, CK~Birdsall, MA~Lieberman, and TD~Rognlien.
\newblock Capacitive rf discharges modelled by particle-in-cell monte carlo
  simulation. i. analysis of numerical techniques.
\newblock {\em Plasma Sources Science and Technology}, 2(4):261, 1993.

\bibitem{kawamura2000physical}
E~Kawamura, Charles~K Birdsall, and Vahid Vahedi.
\newblock Physical and numerical methods of speeding up particle codes and
  paralleling as applied to rf discharges.
\newblock {\em Plasma Sources Science and Technology}, 9(3):413, 2000.

\bibitem{wang2010implicit}
Hong-yu Wang, Wei Jiang, and You-nian Wang.
\newblock Implicit and electrostatic particle-in-cell/monte carlo model in
  two-dimensional and axisymmetric geometry: I. analysis of numerical
  techniques.
\newblock {\em Plasma Sources Science and Technology}, 19(4):045023, 2010.

\bibitem{vahedi1995monte}
Vahid Vahedi and Maheswaran Surendra.
\newblock A monte carlo collision model for the particle-in-cell method:
  applications to argon and oxygen discharges.
\newblock {\em Computer Physics Communications}, 87(1-2):179--198, 1995.

\bibitem{wu2022effects}
Hao Wu, Zhaoyu Chen, Shimin Yu, Qixuan Wang, Xiandi Li, Wei Jiang, and
  Ya~Zhang.
\newblock The effects of match circuit on the breakdown process of capacitively
  coupled plasma driven by radio frequency.
\newblock {\em Journal of Applied Physics}, 131(15), 2022.

\end{thebibliography}

\bibliographystyle{unsrt}

\end{document}